\documentclass[prl,twocolumn,aps]{revtex4}

\usepackage{epsf}

\begin{document}

%\draft

\title{Bouncing localized structures in a Liquid-Crystal-Light-Valve experiment}

\author{M.G. Clerc$\dagger$, A. Petrossian and S. Residori}
\affiliation{Institut Non-Lin\'eaire de Nice, 1361 Route des Lucioles,
06560 Valbonne, France
\\
$\dagger$ Departamento de F\'{\i}sica, Facultad de Ciencias
F\'{\i}sicas y Matem\'aticas, Universidad de Chile,   Casilla
487-3, Santiago, Chile.}

\date{\today}

\begin{abstract}
Experimental evidence of bouncing localized structures in a
nonlinear optical system is reported. Oscillations in the  position of the localized states 
are described by a 
consistent amplitude equation, which we call the
Lifshitz normal form equation, in analogy with phase transitions.
Localized structures are shown to
arise close to the Lifshtiz point, where non-variational terms drive the
dynamics into complex and oscillatory behaviors.
\end{abstract}

\smallskip

%\begin{multicols}{2}
%\narrowtext

\maketitle

Pacs: 05.45.-a,
%Nonlinear dynamics and nonlinear dynamical systems
42.65.Sf,
%Dynamics of nonlinear optical systems; optical instabilities, optical chaos and complexity,
47.54.+r
%Pattern selection; pattern formation}

During the last years localized structures have
been observed in different fields, such as domains in magnetic materials
\cite{Eschenfelder}, chiral bubbles in liquid crystals \cite{Oswald}, current filaments 
in gas discharge
experiments \cite{Astrov97}, spots in chemical reactions \cite{Swinney},
pulses \cite{Wu}, kinks \cite{Denardo} and 
localized 2D states \cite{Edwards} in fluid surface waves, 
oscillons in granular media
\cite{Melo}, isolated states in thermal convection \cite{Heinrichs,Kolodner},
solitary waves in nonlinear optics
\cite{Newell,Oppo,Ackemann} and cavity solitons in lasers \cite{Barland}. 
Localized states are patterns which extend only over a small
portion of a spatially extended and homogeneous system
\cite{Cross}. Different
mechanisms leading to stable localization have been proposed
\cite{Coullet2002}.
Among these, two main classes of localized structures have to be distinguished,
namely those localized structures arising as solutions of a quintic Swift-Hohenberg
like equation \cite{Coullet2002} and those 
that are stabilized by nonvariational terms in the subcritical Ginzburg-Landau
equation \cite{Fauve}. The main difference between the two cases is that 
the first-type localized structures have a characteristic size which is
fixed by the pinning mechanism over the underlying pattern or by spatial damped
oscillations between homogenous states\cite{Coullet2002,CoulletPRL},
whereas the second-type ones have no intrinsic spatial length, their size 
being selected by non-variational effects and going to infinity 
when dissipation goes to zero. In both cases, non-variational effects may lead
to dynamical behaviors of localized structuresÊ\cite{Descalzi}.
Variational models based on a  generalized
Swift-Hohenberg equation have been proposed to describe the appearance of localized 
structures in
nonlinear optics \cite{Lefever}. However, a generalization including
non-variational terms is generically expected to apply even in optics, 
as happens, for instance, in semiconductor laser instabilities \cite{Tlidi}, giving 
rise to dynamical behaviors of localized structures, such as propagation
and oscillations of their positions \cite{Marcel2003}.

We report here an experimental evidence of localized
structures dynamics in a Liquid-Crystal-Light-Valve (LCLV)  with
optical feedback.
It is already known that, in the
simultaneous presence of bistability and pattern forming
diffractive feedback, the LCLV system shows
localized structures \cite{Oppo,Arecchi,Iino,ram}. 
Recently, rotation of localized structures along concentric rings have been
reported in the case of a rotation angle introduced in
the feedback loop \cite{rings}. Here, we fix a zero
rotation angle and we show 
a new dynamical behavior, the bouncing of two adjacent localized structures,
that is not related to imposed boundary conditions but is instead a direct
consequence of the non-variational character of the system under study.
Theoretically, we show that the
LCLV system has several branches of bistability connecting an
homogeneous state to a patterned one and we derive an amplitude
equation accounting for the appearance of localized structures.
This is a one-dimensional model, that we call the Lifshitz normal form
equation \cite{Marcel2003}, characterizing the dynamics of localized structures close to
each point of nascent bistability. 

\begin{figure} [h!]
\centerline{\epsfxsize=7.5 truecm \epsffile{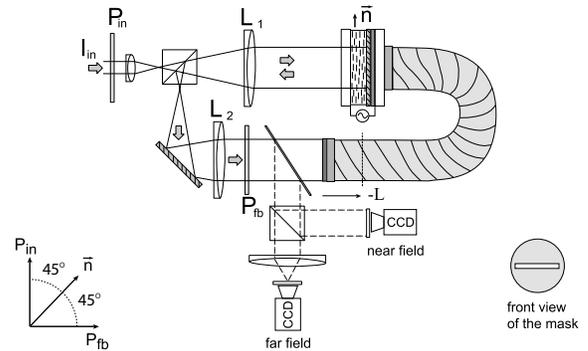}}
\caption{Experimental setup: $\vec n$ liquid crystal
nematic director;  $P_{in}$ and $P_{fb}$ input and
feedback polarizers; $L_1$ and $L_2$ confocal $25$ $cm$
focal length lenses. $-L$ is the free propagation length, negative 
with respect to the plane on which a 1:1 image of the front side of the LCLV
is formed.
\label{setup}}
\end{figure}

{\it Description of the experiment.} The experimental setup is
shown in Fig.\ref{setup}. The LCLV is composed of a nematic liquid
crystal film sandwiched in between a  glass and a photoconductive
plate over which a dielectric mirror is deposed. The liquid
crystal film is planar aligned (nematic director $\vec n$ parallel
to the walls), with a	thickness $d=15$ $\mu m$. The liquid crystal
filling our LCLV is the nematic LC-654, produced by NIOPIK (Moscow) \cite{LC}.
It is a mixture of cyano-biphenyls, with a 
positive dielectric anisotropy $\Delta \varepsilon=\varepsilon_\parallel - \varepsilon_\perp= +10.7$
and large optical birefringence,
$\Delta n=n_\parallel - n_\perp= 0.2$, where $\varepsilon_\parallel$ and $\varepsilon_\perp$ 
are the dielectric permittivities $\parallel $ and
$\perp $ to $\vec{n}$, respectively, and $n_\parallel$ and
are $n_\perp$ are
the extraordinary ($\parallel $ to $\vec{n}$) and ordinary
($\perp $ to $\vec{n}$) refractive index, respectively.
Transparent electrodes over the glass plates permit the application of
an electrical voltage across the liquid crystal layer. 
The photoconductor behaves like a
variable resistance, which decreases for increasing illumination.
The feedback is obtained by sending back onto the photoconductor
the light which has passed through the liquid-crystal layer and
has been reflected by the dielectric mirror. This light beam
experiences a phase shift which depends on the liquid crystal
reorientation and, on its turn, modulates the effective voltage
that locally applies to the liquid crystals.

The feedback loop is closed by an optical fiber bundle  and is
designed in such a way that diffraction and polarization
interference  are simultaneously present \cite{Oppo}. The optical
free propagation length is fixed to $L=-10$ $cm$. At the linear
stage for the pattern formation, a negative propagation distance
selects the first unstable branch of the marginal stability curve,
as for a focusing medium. The angles of the polarizers are at
$45^o$ with respect to the liquid crystal director $\vec n$. The
free end of the fiber bundle is mounted on a precision rotation
and translation stage, to avoid rotation or translation
in the feedback loop. 

\begin{figure} [h!]
\centerline{\epsfxsize=6.2 truecm \epsffile{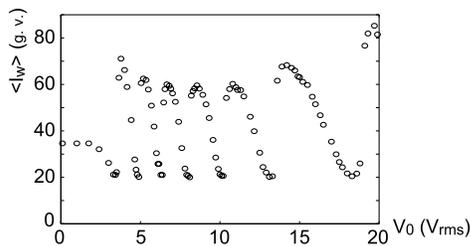}}
\caption{Spatially averaged feedback intensity $<I_w>$ (units are the gray values, g.v., on the CCD camera)
as a function of $V_0$; input intensity
$I_{in}=0.75$ $mW/cm^2$. \label{averageI}}
\end{figure}

For this parameter setting, as $V_0$
increases there is a series of successive branches of bistability
between a periodic pattern and a homogeneous solution. 
In Fig.\ref{averageI} we report the spatially averaged feedback
intensity $<I_w>$ measured for a fixed value of the input intensity, $I_{in}=0.75$
$mW/cm^2$, and for varying $V_0$, by integrating the images on the near-field CCD camera (see Fig.\ref{setup}) . 
The abrupt changes of $<I_w>$ correspond to the 
appearance of localized structures and thus roughly indicate the locations of the nascent 
bistability points. The peak value intensity of the localized structures is approximately
twice the average value $<I_w>$.
In the  LCLV system, the bistability between
homogenous states results from  the subcritical character of the
Fr\'eedericksz transition, when the local electric field, which
applies to the liquid crystals, depends on the liquid crystal
reorientation angle \cite{Clerc,EPJD}. Here, we
limit our study to the bistable branch located around $V_0 = 13.2$
$V_{rms}$ (frequency $5$ $KHz$), however similar observations can be obtained 
close to any other of the nascent bistability points.

We have  carried out one-dimensional experiments, in order to
avoid the influence of any optical misalignment (such as small
drifts) on the dynamics of localized states. A rectangular mask is
introduced in the optical feedback loop, just in contact to the
entrance side of the fiber bundle. The width of the aperture
is $D=0.50$ $mm$ whereas its length is $l=20$ $mm$.
The size of each localized structures is
$\Lambda \simeq 350$ $\mu m$, so that the transverse
aspect ratio $D/ \Lambda \simeq 1$ is small enough for the system 
to be considered as one-dimensional and the longitudinal
aspect ratio $l/ \Lambda \simeq 60$ is large enough for the system to be considered
as a spatially extended one.

\begin{figure} [h!]
\centerline{\epsfxsize=7.0 truecm \epsffile{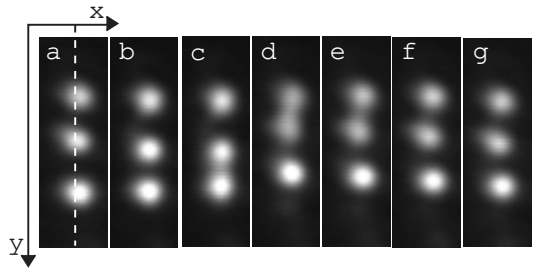}}
\caption{Snapshots showing bouncing  localized
structures; a) $t=0.0$, b) $1.0$, c) $1.3$, d) $1.7$, e)
$2.1$, f) $2.4$, g) $2.8$ $sec$. 
\label{bouncing}}
\end{figure}

\begin{figure} [h!]
\centerline{\epsfxsize=8.5 truecm \epsffile{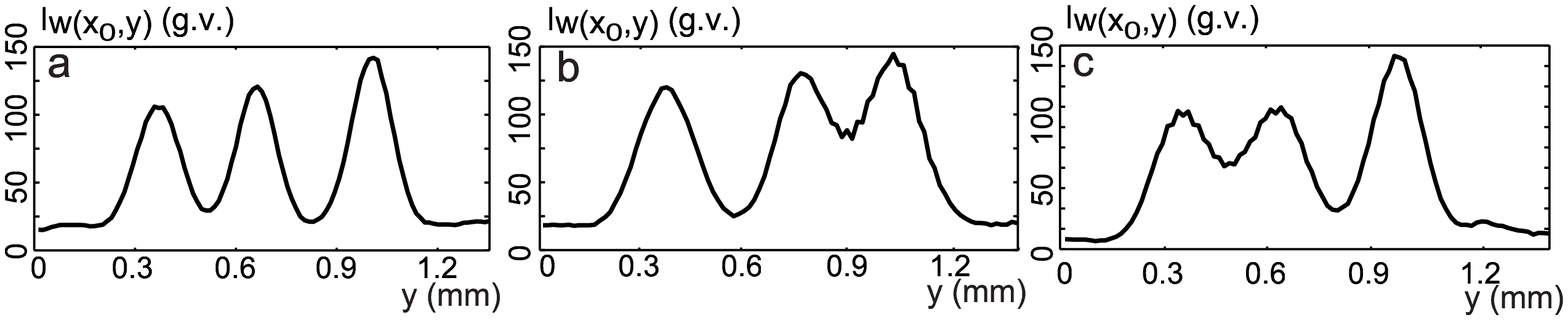}}
\caption{Localized structures profile, $I_w(x_0,y)$; $x_0$ 
is the location of the dashed line in Fig.\ref{bouncing}; a) $t=0.0$
b) $1.3$, c) $1.7$ $sec$. 
\label{prof}}
\end{figure}

\begin{figure} [h!]
\centerline{\epsfxsize=6 truecm \epsffile{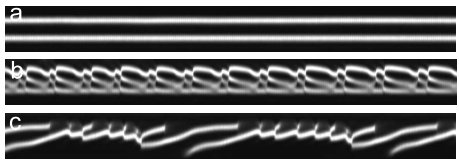}}
\caption{Space(vertical)-time (horizontal) diagrams showing  a)
two stationary localized structures, b) periodic and c) aperiodic oscillations of the structure positions. 
The total elapsed time is $94$ $s$. \label{s-t}}
\end{figure}

Instantaneous snapshots of bouncing localized structures, together with their spatial profiles,
are shown in Fig.\ref{bouncing}, for $V_0=13.2$ $V_{rms}$
and $I_{in}=0.95$ $mW/cm^2$.  
The corresponding 
spatio-temporal plot is displayed in Fig.\ref{s-t}b, showing the
periodic oscillations for the positions of the structures.
In the same figure, Fig.\ref{s-t}a, it is shown the spatio-temporal plot corresponding
to stationary localized structures, as observed for a slightly 
decreased input intensity, $I_{in}=0.90$ $mW/cm^2$, and for the same 
value of $V_0$. 
Fig.\ref{s-t}c displays the spatio-temporal diagram
corresponding to aperiodic oscillations in the structure
positions, as observed for $V_0=13.3$ $V_{rms}$ and $I_{in}=0.90$ $mW/cm^2$.
The dynamical behavior of localized structures is very sensitive 
to parameter changes and, even though their appearance is clearly located around
each point of nascent bistability, their stability range is smaller than the
width of the bistable region. When loosing stability, localized structures
either form clusters or annihilate, depending if they are driven on the pinning
or depinning side of the bistable region \cite{CoulletPRL}. However, a careful
experimental characterization of the pinning front is a work still in progress.

{\it Theoretical description.} The light intensity $I_{w}$
reaching the photoconductor is given by \cite{Oppo},
$I_{w}={I_{in} \over 2}|e^{-i{L \over 2 k} \partial_{xx}} (1+
e^{-i \beta cos^2 \theta})|^{2} \label{eqIPC} $, where $x$ is the
transverse direction of the liquid crystal layer, $\beta cos^2
\theta$ is the overall phase shift experienced by the light
travelling forth and back through the liquid crystal layer; $\beta
= 2kd\Delta {n}$, where $k=2\pi / \lambda $ is the optical wave
number ($\lambda =633$ $nm$).

As long as $I_{in}$ is sufficiently
small, that is, of the order of a few $mW/cm^{2}$, the effective
electric field, $E_{eff}$, applied to the liquid crystal layer can be 
expressed as
$E_{eff}=\Gamma V_0 / d +\alpha I_{w}$, where $V_0$
is the voltage applied to the LCLV, $0< \Gamma < 1$ is a
transfer factor that depends on the electrical impedances of
the photoconductor, dielectric mirror and liquid crystals and $\alpha$ is a
phenomenological dimensional parameter that describes the linear
response of the photoconductor \cite{EPJD}.

Let us call $\theta \left(
x,t\right)$ the average director tilt. $\theta=0$ is the initial  planar
alignment whereas $\theta= \pi /2$ is the homeotropic alignment
corresponding to saturation of the molecular reorientation.
The dynamics  of $\theta$ is  described by a local
relaxation equation of the form

\begin{equation}
\tau \partial _{t}\theta =l^{2}\partial _{xx}\theta -\theta
+\frac{\pi }{2} \left( 1-\sqrt{\frac{\Gamma V_{FT}}{\Gamma{V}_0+\alpha d
I_{w}(\theta,\partial _{x})}}\right) \label{EQ-Valve}
\end{equation}
with $V_{eff} = E_{eff} d = \Gamma{V}_0+\alpha d I_{w}(\theta ,\partial
_{x})>V_{FT}$ the effective voltage applied to the
liquid crystals, $V_{FT}$ the threshold for the Fr\'eedericksz
transition and $l$ the electric coherence length. The above
model have been deduced by fitting the experimental data 
for the open loop response of the LCLV \cite{EPJD}  and it is
slightly different with respect to the one proposed in Ref.
\cite{Oppo}. It is important to note that a rigorous derivation of the 
response function
of the LCLV would require a modal expansion along the longitudinal
direction of the liquid crystal layer.

\begin{figure} [h!]
\centerline{\epsfxsize=5.5 truecm \epsffile{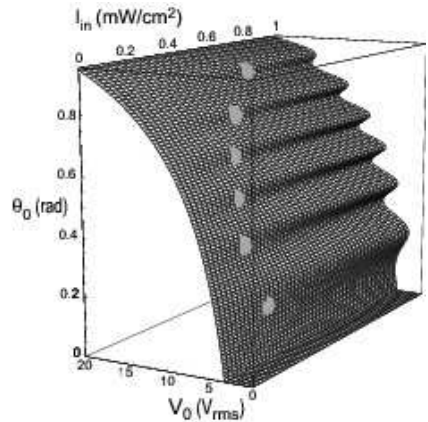}}
\caption{The multi-valued function $\theta_0(V_0,I_{in})$. 
Shaded areas show the location of the nascent bistability points.
 \label{voltage}}
\end{figure}

The homogeneous equilibrium solutions are $\theta_0=0$  when $V_{eff} \le V_{FT}$
and $\theta_0= \pi /2\left( 1-\sqrt{V_{FT}/V_{eff}}\right)$ when $V_{eff} > V_{FT}$. 
Above Fr\'eedericksz transition and by neglecting the spatial terms
we can find a closed expression for the homogeneous equilibrium solutions:
$\theta_0= \pi /2\left( 1-\sqrt{V_{FT}/(\Gamma V_0 + 
\alpha I_{in}[1+\cos(\beta \cos^2 \theta_0)])}\right)$. 
The value of $V_{FT}$ is set to $3.2$ $V_{rms}$, as measured 
for the LCLV \cite{Clerc,EPJD}, and the graph of $\theta_0 (V_0, I_{in})$ is plotted
in Fig.\ref{voltage}.
In agreement
with the bistability branches observed
experimentally, several points of nascent
bistability can be distinguished, corresponding to the
critical points where $\theta_0 ( V_0, I_{in})$
becomes a multi-valued function.

Close to each point of nascent bistability, and
neglecting spatial derivatives, we can develop $\theta=
\theta_0 + u + ...$ and derive a normal form equation describing 
an imperfect pitchfork bifurcation \cite{Cross}, $
\partial _{t}u=\eta +\mu u-u^{3}+h.o.t  \label{EQ-ImpPitchfork}
$, where $\mu $ is the bifurcation parameter and $\eta $ accounts
for the asymmetry between the two homogeneous states. Higher
order terms are ruled out by the scaling analysis, since $u\sim
\mu ^{1/2}$, $\eta \sim \mu ^{3/2}$ and $\partial _{t}\sim \mu $,
$\mu \ll 1$.
 If we now consider the spatial effects, due to the elasticity of
the liquid crystal and to the light diffraction, the system
exhibits a spatial instability as a function of the diffraction
length and, since the spatial dependence of $I_w$ is nonlocal,
the dynamics is a non-variational one.

The confluence of bistability and spatial
bifurcation give rise to a critical point of codimension
three, that we call the Lifshitz point, in analogy with the triple point introduced
for phase transitions in helicoidal ferromagnetic states \cite{Hornreich}. 
Close to this point, we derive an amplitude equation, that we call the Lifshitz normal form
\cite{Marcel2003},

\begin{eqnarray}
\partial _{t}u &=&\eta +\mu u-u^{3}+\nu \, \partial _{xx}u-\partial _{xxxx}u
\nonumber \\
&&+d\text  \, u \, \partial _{xx}u+c \, (\partial _{x}u)^{2},
\label{EQ-Liftchitz}
\end{eqnarray}
where $\partial _{x}\sim \mu ^{1/4}$, $\nu \sim \mu ^{1/2}$
accounts for the intrinsic length of the system (diffusion), $d\sim O\left(
1\right) $ and $c\sim O\left( 1\right) $. 
The term $\partial _{xxxx}u$ describes a 
super-diffusion, accounting for the short distance repulsive interaction,
whereas the terms proportional to $d$ and $c $
are, respectively, the nonlinear diffusion and convection.   
The full and
lengthy expressions of these coefficients, as a function of the
LCLV parameters,
will be reported elsewhere \cite{Rene}. 
Note that the same model has been recently deduced for instabilities in a
semiconductor laser \cite{Tlidi}.

\begin{figure} [h!]
\centerline{\epsfxsize=4 truecm \epsffile{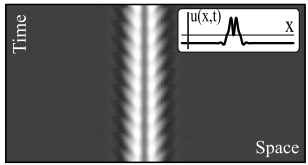}}
\caption{Numerical simulations of Eq. (\ref{EQ-Liftchitz}) for
$\eta = -0.02$, $\mu=-0.02$, $\nu=-1.00$, $c=2.00$ and $d=-1.51$,
showing two bouncing localized
structures (spatial profile in the inset).
\label{numerique}}
\end{figure}

The model shows bistability between a homogeneous and a spatially
periodic solutions and therefore exhibits a family of localized
structures. Depending on the choice of parameters, 
localized structures may show periodic or aperiodic oscillations
of their position. We can fix $\mu$ and $\eta$ by varying $V_0$ and $I_{in}$. 
More interesting is the behavior of the effective diffusion term
$\nu$, which has the form $\nu \propto l^2+(\pi \beta L
\cos^2(\beta/2 \cos^2\theta_o) \sin2\theta_0)/ 4 k ((\Gamma
V_o+\alpha I_{in} (1+\cos(\beta \cos^2\theta_0)))$.
Only when the optical free propagation length $L$ is negative
it is possible, by increasing the input intensity $I_{in}$, 
to drive the system through 
the Lifshitz point 
($\nu$ changes its sign from positive to negative). 
This means that stable localized structures
can be obtained only for a focusing Kerr-like nonlinearity.
Once crossed the Lifshitz point, 
the
nonlinear diffusion coefficient is negative and the convection coefficient
is positive. In this region of parameters, numerical simulations of
Eq.(\ref{EQ-Liftchitz}) 
show a qualitative agreement with the
experimental observations, as shown in Fig.\ref{numerique}.

{\it Conclusions}. We have shown a new kind of localized structure
dynamics, consisting of a bouncing behavior between two adjacent structures,
and we have described it by an universal model, the
Lifshitz normal form equation. The Lifshitz equation,
that reduces to the generalized
Swift-Hohenberg equation for $\eta =d=c=0$, has
already been used to describe the transition from smectic to helicoidal phase 
in liquid crystals \cite{Michelson} and the pulse dynamics in reaction diffusion systems 
\cite{Otha}. When
one neglects the cubic and the nonlinear diffusion terms ($d=0$),
it reduces to the Nikolaevskii equation that describes
longitudinal seismic waves \cite{Tribelsky}.

We gratefully acknowledge Ren\'e Rojas for help in
calculations. The simulation software {\it DimX} is property of INLN. 
This work has been supported by the ACI Jeunes of the French Ministry of
Research (2218 CDR2). M.G. Clerc thanks the support of Programa de
inserci\'{o}n de cient\'{i}ficos Chilenos of Fundaci\'{o}n Andes,
FONDECYT project 1020782, and FONDAP grant 11980002.

%\bigskip
%\bigskip

%\bigskip
%\bigskip

%\bigskip
%\bigskip

%\end{multicols}
\end{document}